\documentstyle[12pt,aaspp4,psfig]{article}

\newcommand{\be}{\begin{equation}}
\newcommand{\ee}{\end{equation}}

\newcommand{\lsim}{\lower 2pt \hbox{$\, \buildrel {\scriptstyle <}\over
         {\scriptstyle \sim}\,$}}

\received{2002 February 6}
\begin{document}
\newcommand{\figureout}[3]{\psfig{figure=#1,width=6in,angle=#2} 
   \figcaption{#3} }
\title{The Multi-Component Nature of the Vela Pulsar Nonthermal X-ray Spectrum}
 \author{Alice K Harding}
\affil{NASA Goddard Space Flight Center, Greenbelt, MD 20771 USA}
\author{Mark S. Strickman}
\affil{Code 7651.2, Naval Research Lab., Washington, DC USA}
\author{Carl Gwinn}
\affil{Dept. of Physics, University of California, Santa Barbara, CA USA}
\author{P. McCulloch and D. Moffet\altaffilmark{1}}
\affil{University of Tasmania, Tasmania, Australia}
\altaffiltext{1}{Current address Furman University, Greenville, SC USA}
\begin{abstract}
We report on our analysis of a 274 ks observation of the Vela 
pulsar with the Rossi X-Ray Timing Explorer (RXTE).  
The double-peaked, pulsed emission at 2 - 30 keV, which we had 
previously detected during a 93 ks observation, is confirmed 
with much improved statistics.  There is now clear evidence, both
in the spectrum and the light curve, that the emission in the RXTE
band is a blend of two separate non-thermal components.  The spectrum of the
harder component connects smoothly with the OSSE, COMPTEL
and EGRET spectrum and the peaks in the light curve are in phase
coincidence with those of the high-energy light curve.  The 
spectrum of the softer component is consistent with an 
extrapolation to the pulsed optical flux, and the second RXTE pulse
is in phase coincidence with the second optical peak.  In addition,
we see a peak in the 2-8 keV RXTE pulse profile at the radio phase.
\end{abstract}

\section{Introduction}

The Vela pulsar (PSR B0833-45) is the strongest $\gamma$-ray source in the 
sky, but it is one of the most difficult pulsars to detect at X-ray energies.
This is in part because it is embedded in a very bright X-ray synchrotron nebula
providing a large unpulsed background, but also because its pulsed X-ray emission 
is comparatively weak.  Its spectral power peaks in the $\gamma$-ray 
band, a characteristic of most other pulsars having detected emission at high 
energies, but it falls off in the hard X-ray band. 

The Vela pulsar is well-known at high $\gamma$-ray energies.  Its light curve as measured by
CGRO/EGRET (Kanbach et al. 1994) consists of two relatively narrow peaks separated by 0.43 in phase.  The first peak lags the radio peak phase by 0.12.  The region between the two
$\gamma$-ray peaks contains a low level of emission.  Observations using COMPTEL and OSSE on CGRO (Strickman et al. 1996 and references therein) show that, while the light curve is similar in the low-energy $\gamma$-ray band to that at higher energies, the spectrum is considerably harder.  The details of this rollover and comparison of light curve behavior are still sketchy due to the relatively poor statistics of the low energy $\gamma$-ray results. Since various models of pulsar high-energy emission make predictions concerning the shape and location of this rollover, and the behavior of the spectrum below the rollover, better measurement of pulsed emission in the X-ray band is important.

The first detection of pulsed emission at X-ray energies was 
made by ROSAT in the 0.1 - 2 keV band (Ogelman et al. 1993).  The observed spectrum
was consistent with a blackbody.  Given the nonthermal nature of pulsed emission at other energies, this component may come from a distinct emission mechanism such as thermal emission from the hot polar cap or neutron star surface.  The first detection of a nonthermal X-ray pulsed component, made with our first RXTE observation,  is reviewed in the next section.

In this paper, we report the results of our analysis of a 300 ks RXTE Cycle 3  
observation.  With significantly improved statistics, we are able to clearly confirm
the existence of at least two non-thermal spectral components and present phase-resolved
spectra.

\section{RXTE Cycle 1 Results}

We detected the Vela pulsar for the first time in the 2 - 30 keV band during a 93 
ks RXTE Cycle 1 observation (Strickman, Harding \& De Jager 1999 [SHD99]) during January 1997.
We found pulsed 
emission with a high degree of significance in each of three broad
energy bands: 2 - 8 keV, 8-16 keV and 16-30 keV.  The light curve in each of these energy
bands showed two peaks separated by about 0.4 in phase, very similar to the light curves
seen by CGRO instruments.  The two peaks, however, exhibited quite different characteristics.
The first peak was very narrow and was phase-aligned with the EGRET $\gamma$-ray
peak.  The second peak was much broader and appeared to shift in phase
(away from the first peak) from lower to higher energies.  
In the 16-30 keV band, this peak was near the phase of the EGRET second peak,
but in the 2-8 keV band the peak was closer to the first RXTE peak and near the phase
of the second optical peak.  Due to limited statistics, we were able to determine the
spectrum only of the two main peaks (Peaks 1 and 2) and what was called the ``Peak 1 
precursor", a broad region of enhanced emission preceding Peak 1.  We found that the
spectrum of Peak 1 (photon index $0.68 \pm 0.14$) was significantly harder than Peak 2  
(photon index $1.17 \pm 0.12$), and both peaks were harder than the Peak 1 precursor
(photon index $1.57 \pm 0.29$).

From these results, we speculated that Peak 2 was a blend of two separate components:
a hard component whose spectrum and light curve connect smoothly with the OSSE, 
COMPTEL and EGRET gamma-ray spectrum and light curves, and a soft component, whose
spectrum and light curve appear to be associated  more with the optical emission.  
We also noted that
the Peak 1 precursor was at the phase of the radio peak and had the softest spectrum
of any RXTE phase components.

\section{RXTE Cycle 3 Observations and Analysis}

The Cycle 3 observations were carried out during April/May and July/August 1998 with a
good exposure of 274 ks. Data were collected in GoodXenon event-by-event mode for the entire observation.  We epoch-folded the data at the pulsar period using the standard fasebin suite of ftools supplied by the RXTE Science Support Center.  The pulsar ephemeris was derived from the same dataset used by the Princeton Pulsar Database (Arzoumanian et al. 1992), but recalculated by D. Nice (private communication, 1999) so as to produce a single ephemeris valid over the entire time span of our observations.  We summed the resulting energy dependent light curves into three broad energy bands (2-8 keV, 8-16 keV and 16-30 keV) to study broad energy dependence of light curve features.

In the analysis of our Cycle 1 observation, we used a simple ``on-pulse'' minus ``off-pulse'' technique to extract phase resolved spectra.  This analysis is inappropriate here since we want to separate spectra from only partially resolved features in the light curve.  Instead we have created a model of the light curve shape based on the broad band light curves.  The model consists of a constant ``background'' level with a set of five sinusoidal ``peaks'' superimposed.  The peaks are each of form
\begin{eqnarray}
C(i) & = & A(i) |\cos[{\pi\over 2}(\phi - \phi_{_0}(i))]|^{\xi(i)}, \\
\nonumber \\
\xi(i) & = & -{0.693\over \ln[\cos({\pi\over 2}W(i))]} \nonumber
\end{eqnarray}
where $A(i)$, 
$\phi_0(i)$ and $W(i)$ are the amplitude, center phase and full width at half maximum of peak $i$
respectively.  

We fit this model to the light curve data in 94 native PCA energy channels roughly spanning the 2-30 keV band.  Preliminary fits to the broad band light curves indicated that, in most cases, $\phi_0(i)$ and $W(i)$ did not vary much with energy.  Hence the values of these parameters were either fixed or computed from  linear functions of energy derived from the preliminary fits.  Hence, the only free parameter in the individual channel light curve fits was $A(i)$.  For each energy channel, the counts for each peak were determined by integrating the appropriate best fit sinusoid over phase.  The resulting spectra were placed in PHA files, where they could be accessed and fit with XSPEC as described in the following section. 

\begin{table}
\caption{Center Phase and Spectrum of Light Curve Peaks} \label{tbl-1}
\begin{center}
\begin{tabular}{lllrr}
\tableline
Peak & $\phi_0$ & Photon index & $\chi^2$ & dof \\
\tableline
1 & $0.117 \pm 0.001$ & $0.8278\pm 0.0928$ & 92.9 & 92\\
2-soft & $0.463 \pm 0.006$ & $2.071 \pm 0.2232$ & 46.8 & 34\\
2-hard & $0.55  \pm 0.008$ (0.50-0.6)& $1.339 \pm 0.1324$ & 53.5 & 62\\
3 & $0.87 \pm 0.02$ & $1.856 \pm 0.528$ & 62.8 & 42\\
4 & $1.006 \pm 0.004$ & $2.057 \pm 0.3136$ & 159.4 & 42\\
\hline
\end{tabular}
\end{center}
\vskip -1.0 truecm
\end{table}

\section{Results}

Figure 1 shows broad-band light curves including the RXTE light curves
in three energy bands together with those of EGRET, OSSE, Chandra and Optical
observations.  The RXTE light curves show a narrow peak (Peak 1) at the phase of the EGRET
first peak and with a phase width of $0.035\pm0.004$, and a second peak that is now clearly seen to be a blend of two
components.  The first of these components, which we call Peak 2-soft, is closer to Peak 1 and is the less intense of the two components.  It is more distinctly visible in the light curve at low energies.  The second component, Peak 2-hard, is further from Peak 1 and is significantly broader than Peak 2-soft, with mean widths in phase of $0.11\pm0.02$ and 
$0.04\pm0.01$ respectively.   There are
two other statistically significant peaks that appear in the RXTE light curve:
Peak 4 at the radio phase (0.0) with width $0.03\pm0.01$ and a weaker peak, Peak 3, 
leading the radio peak with width $0.10\pm0.04$.  These two peaks made up the broad enhancement which we denoted 
the Peak 1 precursor in our Cycle 1 data.  The various peak center phases do not appear to vary with energy except for Peak 2-hard, which is not consistent with a constant.  It varies by 0.08 in phase from the low end to the high end of the RXTE band.  This may in part be an artifact of the sinusoid fitting process.  In all cases, the peak widths are consistent with no variation with energy over the 2-30 keV band.    

To examine phase-resolved spectral behavior, we have
performed fits to the spectra derived from each of the sinusoid model peaks using the full RXTE channel
resolution.  We used the XSPEC WABS*PEGPWRLW spectrum model, a power law with photoelectric
absorption, and have fixed the hydrogen column density to a constant $1 \times 10^{20}\,
\rm atoms\,cm^{-2}$ for all fits, based on ROSAT results (Ogelman et al. 1993).
The resulting best fit photon power-law indices and uncertainties are listed in Table 1, along with $\chi^2$ values.  It is clear that a model in which all the power law indices are consistent with a single value can be rejected handily ($\chi^{2}$ of 32.0 for 4 dof for a model with all indices equal to their mean).  Peak 1 contributes the bulk of $\chi^2$ in this case and is obviously harder than the remainder of the spectra.  The other four spectra are only marginally inconsistent with fluctuations about their mean ($\chi^2$ of 8.7 for 3 dof). 
Figure 2 shows the resulting photon spectra of the different fitted peaks in the
RXTE light curve, along with data from CGRO, ROSAT and optical observations. 
The Peak 1 spectrum seems to connect smoothly with the
OSSE spectrum, confirming and extending the turnover from the COMPTEL and EGRET spectra. The Peak 2-hard spectrum
is much softer than Peak 1 and does not seem to have as smooth a connection to the OSSE spectrum. Peak 2-soft appears softer than Peak 2-hard, but is not well represented by the simple power-law model.  Uncertainties in the two Peak 2 components are increased by possible contamination of each other due to the rather arbitrary nature of our peak shape model and the proximity of the two components.  However, we have no a priori reason to use any specific model or assume a more complex model shape (e.g. asymmetric).  Within these constraints, we believe that these spectra are representative of the component spectra.
The Peak 4 (at the radio phase) spectrum is well resolved but soft and weak.  Within uncertainties, it is indistinguishable from Peak 2-soft.  The Peak 3 (radio precursor) spectrum has not been plotted in Figure 2, 
because the 
errors are large, but it also is consistent with the spectra of Peak 2-soft and Peak 4. 

Our finding of clearly separate emission components in the phase resolved
spectrum is confirmed by examination of the broad-band light curves in Figure 1.  The hardest
RXTE spectral components, Peak 1 and Peak 2-hard, are phase-aligned with EGRET Peaks 1
and 2.  Interestingly, we find no evidence for the interpeak emission component seen
by EGRET.  This is consistent with EGRET's finding that this is the hardest
emission component in the Vela phase-resolved spectrum.  The softest RXTE spectral
components, Peak 2-soft, Peak 3 and Peak 4 are all phase-aligned with features
in the optical and radio light curves.  Peak 2-soft is aligned with the second 
optical peak.  Peak 4, which is strongest in the 2-5 keV band, is at the phase of the 
radio pulse (phase 0.0/1.0) and with a peak in the optical light curve at the radio phase.
Peaks 1 and 2 are roughly in phase with the first two peaks in the Chandra HRC profile.
Peak 3 is in the vicinity of the very strong third peak in the Chandra profile. 

In summary, then, we have identified separate hard and soft emission components in the Vela pulsar spectrum.
An extrapolation of the soft component spectrum to lower energies is roughly
consistent with the optical spectrum (Nasuti et al. 1997, Mignani and Caraveo 2001), 
while the hard component continues trends from higher energy. 
The hardest component, Peak 1, is phase aligned with features at higher energy, while the soft components, Peaks 2-soft and the radio phase peak, are aligned with features at lower 
energies.  Peak 2-hard is an exception, being not much harder than the soft components but 
phase aligning with a $\gamma$-ray feature.  Further observations with improved statistics would allow us to separate the two components of Peak 2 with more confidence.

In Figure 3, we plot our phase-averaged RXTE spectrum for better comparison with
the other phase-averaged results, including the recent Chandra observations (Pavlov
et al. 2001).  Although it seems to smoothly 
bridge the gap between the ROSAT/Chandra and OSSE spectra, the RXTE phase-averaged
spectrum is clearly not a simple power law but a transition between hard and soft
spectral components.  Pavlov et al. (2001) determined that a two-component model,
blackbody plus power law, best fit the Chandra phase-averaged point-source spectrum
in the 0.25 - 8 keV band.  Our phase-averaged spectrum roughly agrees with the power-law component of the Chandra spectrum in hardness, even though it is not well-represented 
by a single power-law.  Also, the current Chandra result represents total flux from the 
point source, while our result is ``on-pulse'' minus ``off-pulse,'' which explains the difference in normalization and could lead to a variation in shape as well.
We will be able to make a better comparison when Chandra phase-resolved spectra are
available.

\section{Discussion}

The RXTE Cycle 3 observations have independently confirmed the multicomponent 
nature of pulsed emission from the Vela pulsar in the energy range 2 - 30 keV as suggested 
by the Cycle 1 observations.  With the improved statistics of the Cycle 3 data, we 
have been able to separate the broad second peak in the RXTE X-ray light curve into 
soft (Peak 2-soft) and hard (Peak 2-hard) spectral components which maintain their phase integrity throughout the RXTE energy range. In addition, we have discovered a new 
feature in
the RXTE light curve: a peak (Peak 4) at the phase of the radio pulse with an extremely 
soft spectrum.  There is, in addition, significant emission leading the radio phase 
(Peak 3).  Peaks 1 and 2-hard make up the hard spectral component 
whose light curve peaks are in phase with those of the gamma-ray 
light curve, and whose spectrum smoothly connects to the 100 keV - 5 GeV spectrum.
The Vela pulsar now joins the Crab pulsar in having phase-aligned pulse profiles
and a continuous spectrum from hard X-ray to hard-gamma rays.  Unlike Vela,
the phase alignment of the Crab pulse profile continues down to the radio band.
RXTE Peaks 2-soft, 3 and 4 make up the soft component and each are phase aligned
with features in optical and radio light curves.  Their spectra, consistent 
with each other in both flux and spectral index, extrapolate to the optical flux points.

The hard X-ray band is a complex transition region in the Vela pulsar spectrum, where at
least two non-thermal emission components possibly produced at several different
locations in the magnetosphere are simultaneously present.  The 2-30 keV Vela spectrum 
may thus be a ``Rosetta Stone" that will provide some valuable clues to
deciphering pulsar emission mechanisms.  Unfortunately, from the RXTE observations
alone the theoretical picture is not yet clear, but there are a few interesting 
connections to be made with the existing models.  

Pulsar high-energy emission models generally have fallen into two main categories.
Polar cap models (e.g. Daugherty \& Harding 1996, Sturner et al. 1995) advocate 
particle acceleration and emission in the magnetic polar
regions within several stellar radii of the neutron star surface.  High energy
photon spectra are produced by primary particle curvature or inverse-Compton
emission and additional synchrotron and inverse-Compton components from  
electron-positron pair cascades.  In outer gap models (e.g. Cheng, Ho \& Ruderman 1986,
Romani 1996), particle acceleration takes
place in the outer magnetosphere, in vacuum gaps along null charge surfaces.
Although the same radiation processes that contribute to polar cap emission also 
produce the outer-gap high-energy emission, the pair production now requires soft
X-ray photons from a hot neutron star rather than a strong magnetic field.
 
Although the RXTE hard component spectrum connects to the CGRO $\gamma$-ray spectrum, 
the X-ray spectrum is harder, requiring a break around 50-100 keV, a feature already
suggested by OSSE (Strickman et al. 1996).  Polar cap models predict such a break 
in the synchrotron spectrum of the pair cascade at the local cyclotron energy, blueshifted 
by the relativistic motion of the pairs along the magnetic field (Harding \& Daugherty 
1999).  The predicted turnover energy is an increasing function of local
field strength and thus provides a measure of the height of the emission above the
neutron star surface.  With pair Lorentz factors around $\Gamma_{\pm} \sim 20$,
the turnover at 50-100 keV in the Vela hard X-ray spectrum would indicate an emission
height of 1-2 stellar radii above the surface.  The outer gap model of Romani (1996)
predicts a turnover in the gamma-ray curvature spectrum and a transition to synchrotron
radiation in the Vela spectrum, but at an energy around 1 MeV.  

The RXTE soft component suggests continuous emission from optical to hard X-rays. 
In polar cap models, this emission cannot be synchrotron radiation since it is
well below the cyclotron energy, but inverse Compton scattering radiation would be
possible.  Zhang \& Harding (2000) have argued that inverse-Compton scattering of
thermal X-rays by cascade pairs will produce observable soft X-ray emission and
the computed spectrum of this radiation (Dyks et al. 2001) is possibly consistent
with the Vela RXTE phase-averaged spectrum and is likely to contribute at least
some of the soft emission component.  However, scattering of much softer photons,
possibly radio, is required to produce a non-thermal emission component extending down 
to the optical band.  The fact that one of the RXTE soft component peaks is at the 
radio phase makes this a potentially attractive prospect for further study.
In outer gap models, synchrotron radiation of backflowing particles from the gap
accelerator produces non-thermal X-rays (Cheng \& Zhang 1999).  The predicted spectral 
index of -1.9 is consistent with the spectral index of the RXTE soft component.

We would like to thank George Pavlov and Divas Sanwal for providing their data on 
the Chandra pulse profile in advance of publication and for comments on the 
manuscript.   We also thank David Nice for providing the ephemeris from the Princeton
Pulsar Database.

\clearpage
\figureout{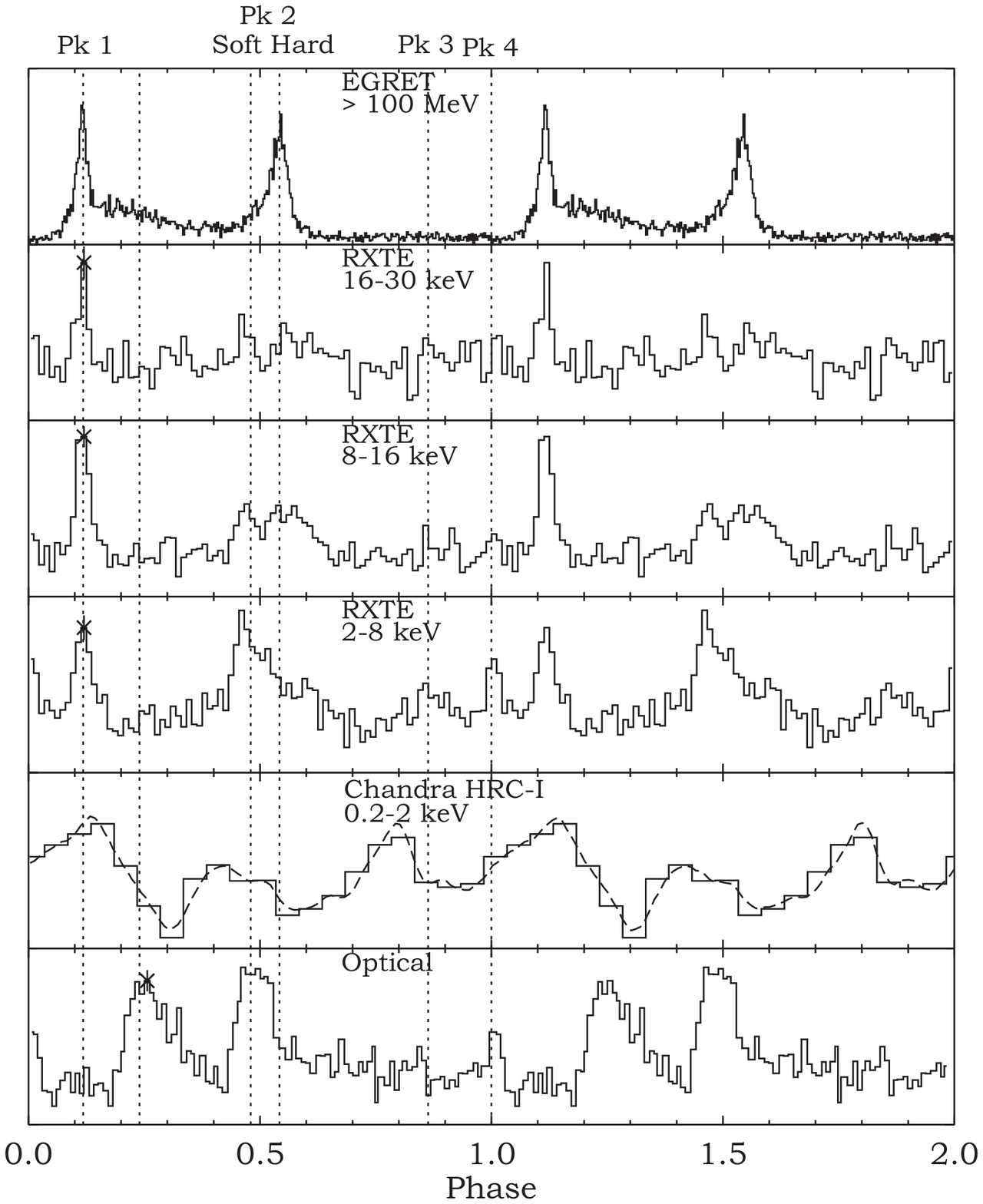}{0}
{Vela Cycle 3 RXTE pulsed emission phase histograms in three broad energy bands,
averaged over the entire observation.  Also shown are pulse profiles in EGRET 
(Kanbach et al. 1994), Chandra (Sanwal et al. 2002) and optical 
(Gouiffes 1998) bands.}

\figureout{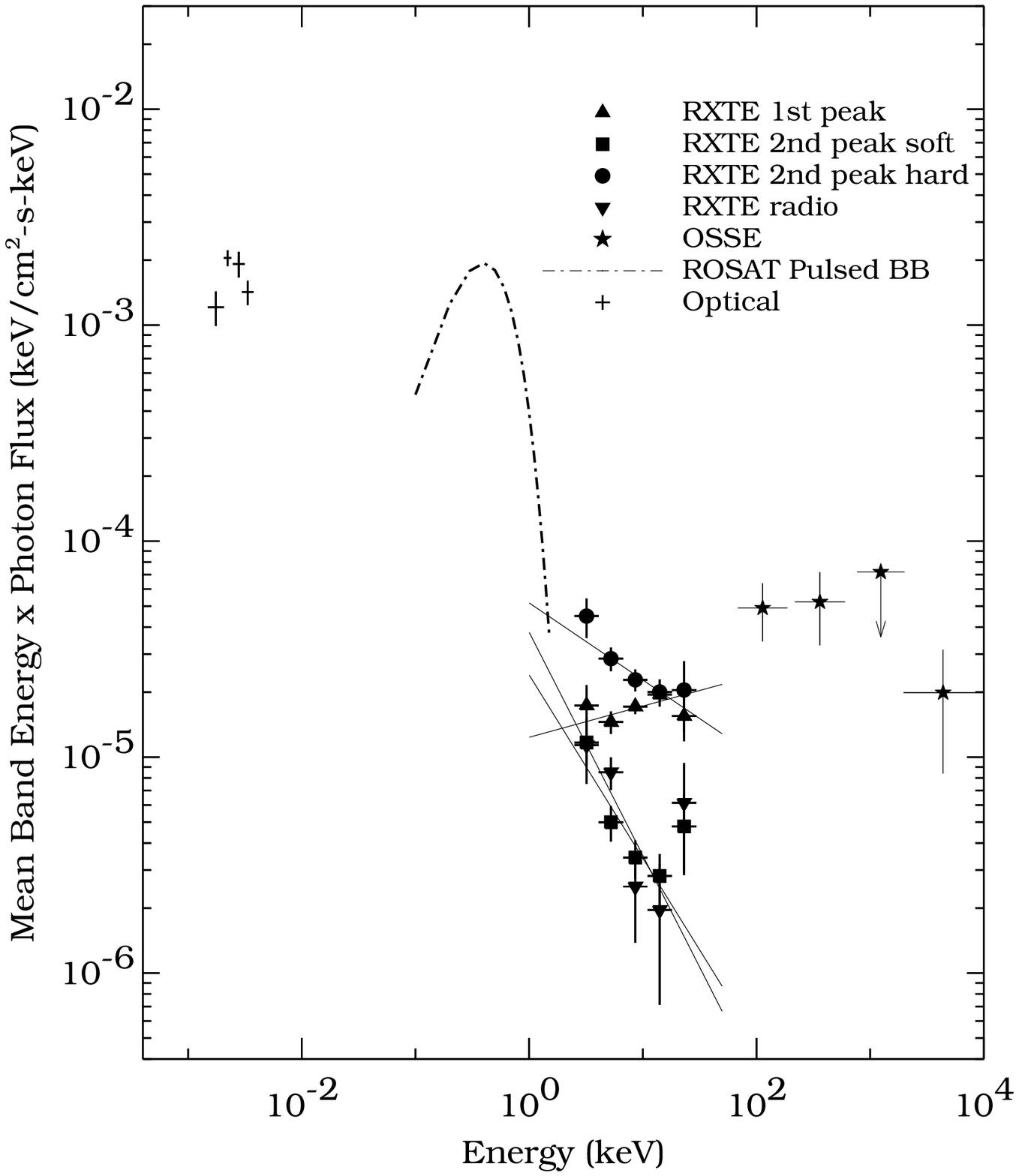}{0}
{RXTE phase-resolved spectra for four peaks in the light curve which have
been fitted with a five-component sinusoidal model (see text), together with OSSE
(Strickman et al. 1996), ROSAT (Ogelman et al. 1993) and optical (Magnani et al. 2001)
phase-averaged spectra.}

\figureout{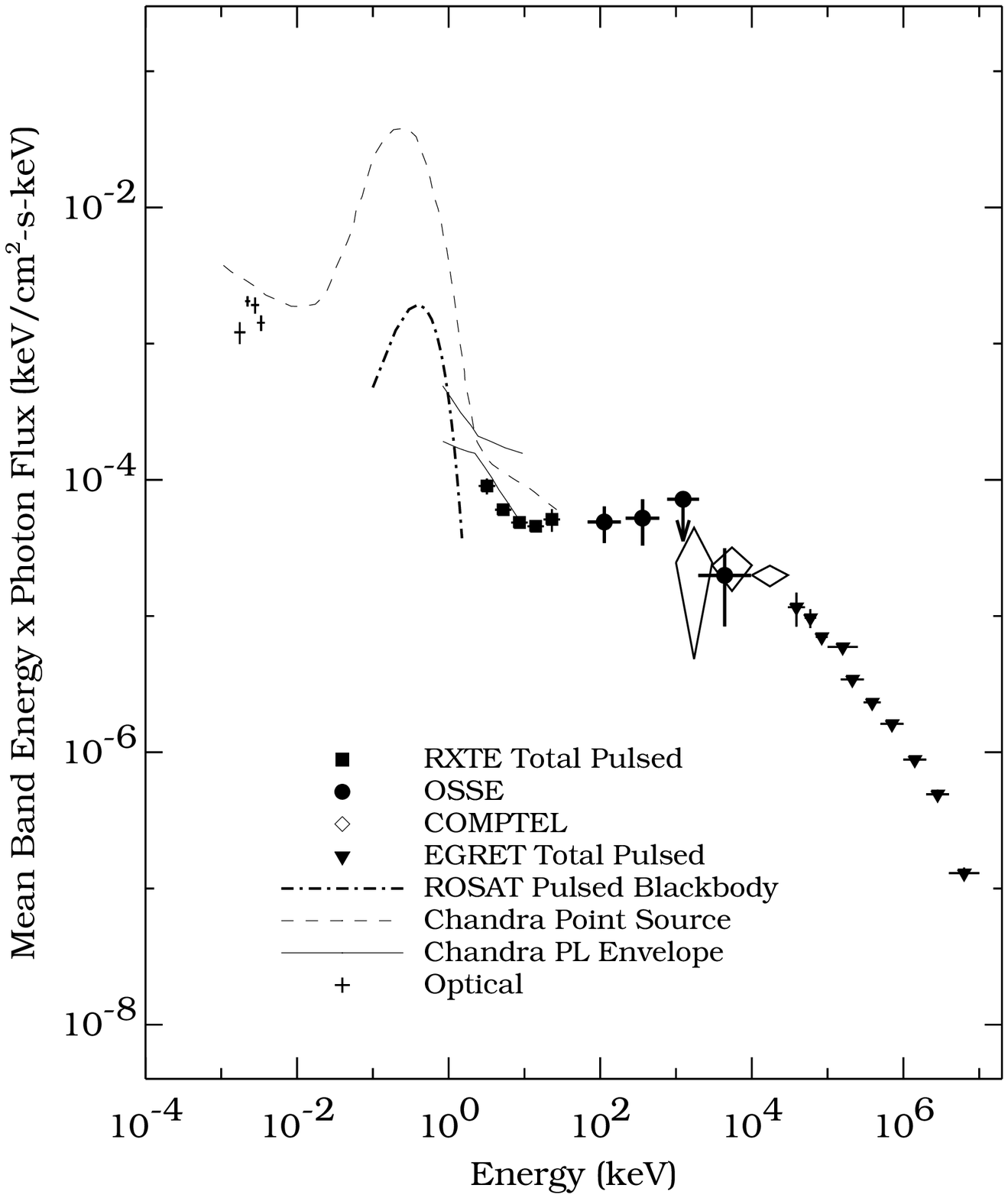}{0}
{Phase-averaged spectra in RXTE, optical (Magnani et al. 2001), Chandra
(Pavlov et al. 2001), OSSE (Strickman et al. 1996), COMPTEL (Schonfelder et al. 1994) 
and EGRET (Kanbach et al. 1994) bands.  The Chandra thermal and power law components 
are plotted separately.}


\begin{references}

\reference{} Arzoumanian, Z., Nice, D. \& Taylor, J. H. \\1992, GRO/radio timing data base,
   Princeton Univ.
\reference{}Cheng, K.~S., Ho, C., \& Ruderman, M.~A. 1986, ApJ, 300, 500. 
\reference{} Cheng, K. S. \& Zhang, L. 1999, ApJ, 515, 337.
\reference {daugh96} Daugherty,~J.~K. \& Harding,~A.~K. 1996, ApJ, 458, 278.
\reference{} Dyks. J., Rudak, B. \& Bulik, T. 2001, Proceedings of the 4th INTEGRAL 
   Workshop, ed. A. Gimenez, V. Reglero \& C. Winkler (ESA SP-459, Noordwijk), p. 191. 
\reference{} Gouiffes, C. 1998, in Neutron Stars and Pulsars, ed. N. Shibazaki, N. Kawai,
   S. Shibata \& T. Kifune (Univ. Acad. Press: Toyko), p. 363.
\reference{} Harding, A. K. \& Daugherty, J. K. 1999, Proc. of 3rd Integral 
Workshop, Astr. Lett. Comm., 38, 25.
\reference{} Kanbach, G.  et al. 1994, A \& A, 289, 855. 
\reference{} Mignani, R. P. \& Caraveo, P. A. 2001, A \& A, 376, 213. 
\reference{} Nasuti, F. P. et al. 1997, A \& A, 323, 839.
\reference{} Ogelman, H., Finley, J.P., and Zimmermann, H.U., 1993, Nature, 361,136.
\reference{} Pavlov, G. G., Zavlin, V. E., Sanwai, D., Burwitz, V. \& Garmire, G. P.
   2001, ApJ, 552, L129.
\reference{} Romani, R.~W. 1996, ApJ, 470, 469.
\reference {stric96} Strickman,~M.S., Grove,~J.E., Johnson,~W.N.,
Kinzer,~R.L., Kroeger,~R.A., Kurfess,~J.D., Grabelsky,~D.A., Matz,~S.M.,
Purcell,~W.R. \& Ulmer,~M.P. 1996, ApJ, 460, 735.
\reference{} Sanwal, D. et al. 2002, in "Neutron Stars in Supernova Remnants" 
(ASP Conference Proceedings), eds P. O. Slane and B. M. Gaensler, in press. 
\reference{} Strickman, M. S., Harding, A. K. \& DeJager, O. C. 1999, ApJ, 524. 373 [SHD99].
\reference{} Sturner, S. J., Dermer, C.~D. \& Michel, F.~C. 1995, ApJ, 445, 736.
\reference {thomp97} Thompson,~D.J., Harding,~A.K., Hermsen,~W. \& Ulmer,~M.P.
1997 in Proceedings of the Fourth Compton Symposium, ed. Charles D. Dermer et
al. (Woodbury, New York:  American Institute of Physics), 39.
\reference{} Zhang, B. \& Harding, A. K. 2000, ApJ, 532, 1150.

\end{references}
\end{document}